\documentclass[useAMS,usenatbib]{mn2e}
\usepackage{txfonts}
\usepackage{longtable}  
\usepackage{rotating}
\usepackage{natbib}
\usepackage{graphicx}
\usepackage{graphics}
\usepackage{psfrag}
\usepackage{amssymb}
\usepackage{multicol}
\usepackage{lscape}
\usepackage{multirow}

\usepackage[usenames]{color}

\bibpunct{(}{)}{;}{a}{}{,}

\def\logTeff{\ensuremath{\log T_{\mathrm{eff}}}}

\def\vsini{\ensuremath{{\upsilon}\sin i}}

\def\logl{$\log L/{\rm L}_{\odot}$}

\def\errlogTeff{$\sigma_{\log T_{\mathrm{eff}}}$}

\def\errlogl{$\sigma_{\log L/{\rm L}_{\odot}}$}

\title[]{$\lambda$ Bootis stars in the SuperWASP survey} 

\author[E. Paunzen et al.]{E. Paunzen,$^{1}$\thanks{epaunzen@physics.muni.cz}
			   M. Skarka,$^{1}$
				 P. Walczak,$^{2,3}$
				 D.L. Holdsworth,$^{4}$
         B. Smalley,$^{4}$
			   \newauthor R.G. West,$^{5}$ and
				 J. Jan{\'i}k$^{1}$ \\
$^{1}$Department of Theoretical Physics and Astrophysics, Masaryk University,
Kotl\'a\v{r}sk\'a 2, 611\,37 Brno, Czech Republic\\
$^{2}$Instytut Astronomiczny Uniwersytet Wroc{\l}awski, Wroc{\l}aw, Poland\\
$^{3}$Nicolaus Copernicus Astronomical Center, Warsaw, Poland\\
$^{4}$Astrophysics Group, Keele University, Staffordshire, ST5 5BG, UK\\
$^{5}$Department of Physics, University of Warwick, Coventry, CV4 7AL, UK}
\begin{document}

\date{}

\pagerange{\pageref{firstpage}--\pageref{lastpage}} \pubyear{2015}

\maketitle

\label{firstpage}

\begin{abstract}
We have analysed around 170\,000 individual photometric WASP measurements of 15 well
established $\lambda$ Bootis stars to search for variability. The $\lambda$ Bootis 
stars are a small group of late-B to early-F, Pop. I,
stars that show moderate to extreme (surface) underabundances (up to a factor
100) of most Fe-peak elements, but solar abundances of lighter elements (C, N,
O and S). They are 
excellent laboratories for the study of fundamental astrophysical processes such as 
diffusion, meridional circulation, stellar winds, and accretion in the presence of 
pulsation. From the 15 targets, eight are variable and seven are apparently constant
with upper limits between 0.8 and 3.0\,mmag.
We present a detailed time series analysis and a comparison with previously published
results. From an asteroseismologic study we conclude that the found
chemical peculiarities are most probably restricted to the surface.
\end{abstract}


\section{Introduction}\label{introduction}

The group of $\lambda$\,Bootis stars stand out among the chemically peculiar (CP) stars of
the upper main-sequence. The CP stars normally exhibit strong overabundances
of elements, probably caused by magnetic fields, slow rotation or
atmospheric diffusion \citep[e.g., ][]{Neto08}. However, the $\lambda$\,Bootis stars
are a small group (only 2\,per cent) of late-B to early-F
stars that show moderate to extreme (surface) underabundances (up to a factor
100) of most Fe-peak elements, but solar abundances of lighter elements (C, N,
O and S). Several members of the group
exhibit a strong infrared excess, and a disk \citep{Paun03,Boot13}.

To explain the peculiar chemical abundances, \citet{Venn90} suggested they are
caused by selective accretion of circumstellar material. One of the principal 
features of that hypothesis is that the observed
abundance anomalies are restricted to the stellar surface. On the basis of this hypothesis \citet{Kamp02} and
\citet{Mart09} developed models which describe the interaction of the star with
its local interstellar and/or circumstellar environment, whereby different
degrees of underabundance are produced by different amounts of accreted
material relative to the photospheric mass. The fact that the fraction of
$\lambda$\,Bootis stars on the main-sequence is so small would then be a
consequence of the low probability of a star-cloud interaction within a limited
parameter space. For example, the
effects of meridional circulation dissolves any accretion pattern a few
million years after the accretion has stopped.

Since the early 1990s, the pulsational behaviour of the group members was extensively studied because
almost all $\lambda$\,Bootis stars are located within the classical $\delta$\,Scuti/$\gamma$\,Doradus 
instability strip \citep{Paun02a,Breg06,Paun08,Murp14}.
It was deduced that at least 70\,per cent of the group members inside the classical instability strip pulsate. 
They do so with first and second overtones modes (Q\,$<$\,0.020\,d) typical for $\delta$\,Scuti type pulsators. 
Only a few stars, if any, pulsate in the fundamental mode. In general, the amplitudes do not exceed a few mmags.
The Period-Luminosity-Colour relation for this group is, within the errors, identical to that of the normal 
$\delta$\,Scuti stars \citep{Paun02a}.

In this paper we have selected fifteen targets from the 66 $\lambda$\,Bootis stars in the lists of \citet{Gray98} and \citet{Paun01a} 
which have at least 1000 data points in the Wide Angle Search for Planets (WASP) archive and are fainter than $V$\,=\,8\,mag to 
avoid the effects of saturation. We analysed the time series of these fifteen group members to search for variability 
and compared our result with those in the literature.

Observations, target selection, and data 
analysis are described in Sect.~\ref{selection}; results are presented and discussed in Sect.~\ref{analysis}. We 
conclude in Sect.~\ref{conclusion}.

\begin{table*}
\caption{Astrophysical parameters of our fifteen target stars together with previous published results
about their pulsational behaviour. For constant
stars the Amp column gives the upper limit to pulsational amplitudes.}
\label{targets_basics}
\begin{tabular}{ccccccccccccc}
\hline\hline \\ 
HD & HIP/DM/TYC & $V$   & \logTeff & \errlogTeff & \logl & \errlogl & Const/Var & $\Delta t$ & Filter & $f$   & Amp & Ref \\     
   &        & [mag] & [dex]    & [dex]       & [dex] & [dex] &   &  [h]       &        & [d$^{-1}$] & [mmag] & \\   
\hline
23392	&	17462	&	8.26	&	3.991	&	0.012	&	1.40 & 0.08 & C & 9.5 & $b$ & $-$ & 2.0 & C \\
36726	&	BD$-$00~993	&	8.84	&	3.978	&	0.010	&	1.36 & 0.07 & C & 9.3 & $b$ & $-$ & 3.0 & C\\ 
83041	&	47018	&	8.93	&	3.852	&	0.013	&	1.24 & 0.07 & V & 9.3 & $b$ & 15.16 & 7.0 & B \\
83277	&	47155	&	8.30	&	3.845	&	0.012	&	1.31 & 0.09 & C & 3.5 & $b$ & $-$ & 1.4 & D \\
90821	&	BD+26~2097	&	9.47	&	3.913	&	0.004	&	1.60 & 0.10 & C & 2.0 & $b$ & $-$ & 2.2 & D \\
101108	&	56768	&	8.88	&	3.893	&	0.004	&	1.29 & 0.08 & C & 24.5 & $b$ & $-$ & 2.0 & C \\
105058	&	58992	&	8.90	&	3.889	&	0.010	&	1.40 & 0.09 & V & 1.8 & $b$ & 24.83 & 3.0 & C \\
120896	&	67705	&	8.49	&	3.861	&	0.009	&	1.09 & 0.08 & V & 3.9 & $b$ & 17.79 & 10.0 & D \\
125889	&	CD$-$35~9471	&	9.85	&	3.862	&	0.010	&	0.91 & 0.09 & & & & & & \\
184779	&	CD$-$44~13449	&	8.94	&	3.858	&	0.010	&	1.25 & 0.06 & & & & & & \\
191850	&	CD$-$46~13448	&	9.65	&	3.869	&	0.009	&	1.19 & 0.07 & V & 7.1 & $b$ & 13.53 & 34.0 & A \\
261904	&	0750$-$1747$-$1	&	10.21	&	3.974	&	0.009	&	1.48 & 0.12 & C & 4.3 & $b$ & $-$ & 3.5 & D \\
290492	&	BD$-$00~980	&	9.31	&	3.908	&	0.008	&	1.02 & 0.11 & C & 9.2 & $b$ & $-$ & 1.8 & D \\
290799	&	BD$-$00~1047 &	10.70	&	3.889	&	0.005	&	0.83 & 0.12 & V & 7.8 & $b$ & 23.53 & 6.0 & D \\
294253	&	BD$-$03~1154 	&	9.66	&	4.027	&	0.007	&	1.58 & 0.11 & C & 4.7 & $V$ & $-$ & 7.0 & C \\
\hline \\  
\multicolumn{11}{l}{A ... \citet{Paun95}; B ... \citet{Paun96}; C ... \citet{Paun98}; D ... \citet{Paun02a}} \\
\end{tabular}                                          
\end{table*}

%
\begin{figure}
\begin{center}
\includegraphics[width=85mm,clip]{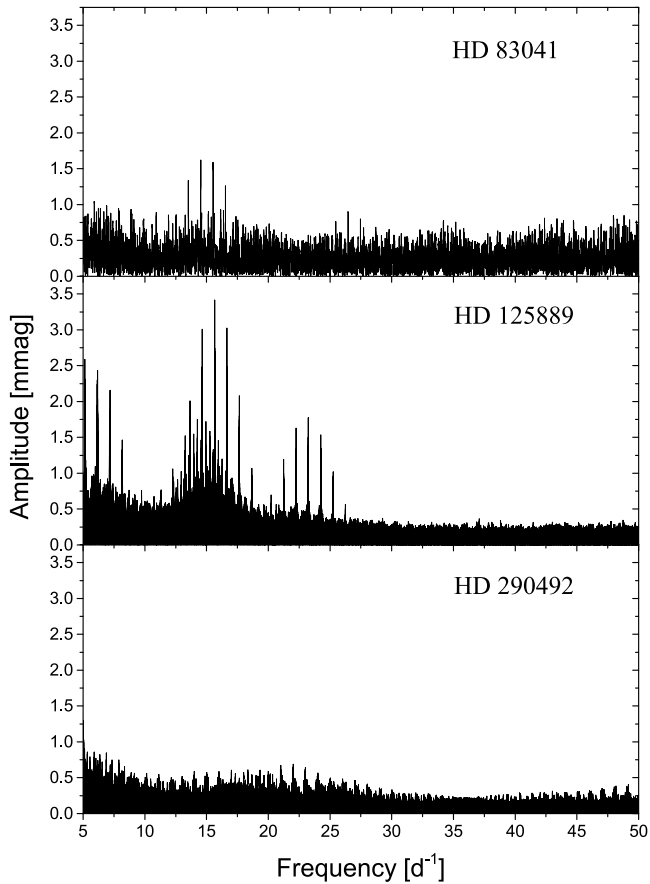}
\caption{Fourier spectra of HD\,83041 (single frequency), HD\,125889 (multiperiodic), and
HD\,294092 (no significant frequency detected).}
\label{fourier} 
\end{center} 
\end{figure}

%
\begin{figure}
\begin{center}
\includegraphics[width=85mm,clip]{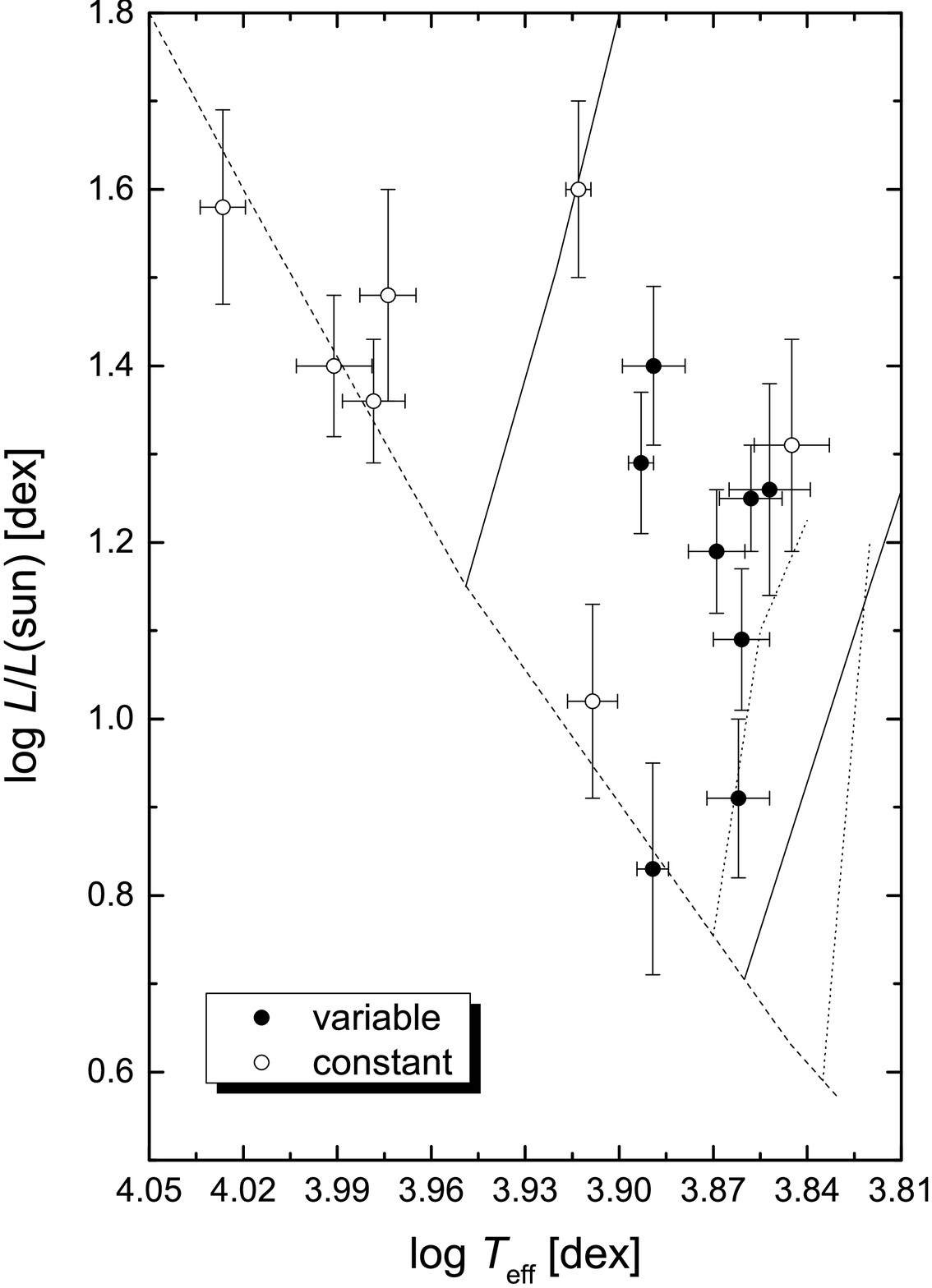}
\caption{The \logTeff\ versus \logl\ diagram of our programme stars (Table \ref{targets_basics}). The borders
of the $\delta$\,Scuti (solid lines) and $\gamma$\,Doradus (dotted lines) instability strips taken from \citet{Breg98} and \citet{Dupr04},
respectively. The zero-age main-sequence (dashed line) is taken from \citet{Clar95}.}
\label{instability} 
\end{center} 
\end{figure}

%
\section{Target selection, observations and reductions} \label{selection}

The photometric data used for this study was from SuperWASP; the WASP instruments are described in \citet{Poll06}, with the reduction techniques described in
\citet{Smal11} and \citet{Hold14}. The aperture-extracted photometry from each camera on each night are corrected for primary
and secondary extinction, instrumental colour response and system zero-point relative to a network of local secondary 
standards. The resultant pseudo-$V$ magnitudes are comparable to Tycho $V$ magnitudes.

We have selected all 66 bona fide $\lambda$\,Bootis stars from the lists by \citet{Gray98} and \citet{Paun01a} which are fainter than
8th magnitude to avoid effects of saturation of the photometric data. Furthermore, only objects with more than 1000
data points for which no two approximately equal brightness stars within the 3.5-pixel
($\sim$50{\arcsec}) WASP photometry aperture were processed. In total, we analysed the light curves of fifteen group members.

For two stars, HD\,101108 and HD\,290492, the analysed WASP photometry includes a companion. BD+39\,2458B, the companion of HD
101108, is about 7{\arcsec} away and 2.6 magnitude fainter, while the companion for HD\,290492, TYC\,4766-2124-1 \citep[spectral type of G8\,III, ][]{Paun02b}
is at a distance of about 25{\arcsec} and $\Delta V$\,=\,2.0\,mag.

Employing the methods and starting values listed by \citet{Paun02a,Paun02b} we re-evaluated the \logTeff\, and \logl\, for
our targets. Photometric data were taken from the General Catalogue of Photometric Data 
\citep[GCPD, ][]{Merm97}\footnote{http://gcpd.physics.muni.cz/}.
Where possible, averaged and
weighted mean values were used throughout. Hipparcos parallax measurements \citep{Leeu07} for stars with a precision better than 30\%
were used. Table \ref{targets_basics} lists the basic information, the \logTeff\, and \logl\, together with results from former pulsation studies.

The light curves were examined in more detail using the {\sc period04} program \citep{Lenz05}, which performs a 
discrete Fourier transformation. 
The results were checked with those from the {\sc cleanest} and Phase-Dispersion-Method computed 
within the programme package Peranso\footnote{http://www.peranso.com/} \citep{Husa06}. The differences for the different
methods are within the derived errors depending on the time series characteristics, i.e. the distribution of the measurements over
time and the photon noise. We applied these methods to the data sets of four pulsating metallic-lined Am stars (Renson 1984, 29800, 37494, and
55094) taken from \citet{Smal11}.
The errors are in the same range as for our target stars lending confidence in the reduction and analysis techniques.

The detailed observational dates and results of the time series analysis for all targets are listed in 
Table \ref{results_data}.

\begin{table*}
\centering
\caption{Data characteristics, frequencies, amplitudes, and upper limits for our targets. The results of the individual stars are described in more detail in Sect.\,\ref{analysis}.}
\label{results_data}
\begin{tabular}{rcrrrrc}
\hline\hline \\ 
\multicolumn{1}{c}{HD} & HJD(start) & \multicolumn{1}{c}{$\Delta t$} & \multicolumn{1}{c}{$N_{\mathrm{data}}$} & \multicolumn{1}{c}{Frequency} & \multicolumn{1}{c}{Amplitude} & 
\multicolumn{1}{c}{Upper limit} \\
& [2450000+] & \multicolumn{1}{c}{[d]} & & \multicolumn{1}{c}{[d$^{-1}$]} & \multicolumn{1}{c}{[mmag]} & \multicolumn{1}{c}{[mmag]} \\
\hline
23392	&	4721.63574 & 859.86426 & 14887 & & & 1.0 \\
36726	&	5496.45215 & 478.95117 & 5232 & & & 1.5 \\ 
83041	&	3860.20096 & 2204.16476 & 11178 & 14.5293 & 1.6 \\
83277	&	3860.23315 & 749.11744 & 5377 & & & 3.0$^a$ \\
90821	&	3131.37378 & 1094.15601 & 4096 & & & 0.8 \\
101108	&	3128.40356 & 1108.13257 & 4579 & 26.6100 & 2.2 &     \\
105058  &	5651.40381 & 56.058100 & 9137  & 19.8097 & 8.9 &     \\
        &                  &           &       & 8.9567  & 4.5 &     \\
        &                  &           &       & 9.1293  & 3.9 &     \\
        &                  &           &       & 10.2565 & 3.5 &     \\
        &                  &           &       & 17.7170 & 3.1 &     \\
        &                  &           &       & 20.7920 & 3.3 &     \\
        &                  &           &       & 14.1886 & 2.2 &     \\
        &                  &           &       & 15.9500 & 2.7 &     \\
120896	&	4516.72363 & 1100.87598 & 4443 & 7.6138 & 5.6 \\
& & & & 19.6300 & 3.5 \\
& & & & 12.2889 & 3.0 \\
& & & & 8.4705 & 2.6 \\
& & & & 9.9511 & 2.4 \\
& & & & 21.0899 & 2.3 \\
& & & & 3.0111 & 2.2 \\
& & & & 20.1347 & 2.0 \\
125889	&	3860.38989 & 2246.07593 & 28492 & 15.6547 & 3.5 \\
        &                  &           &       & 5.1457  & 2.7     \\
        &                  &           &       & 6.1680  & 2.1     \\
184779	&	3860.44019 & 2246.95336 & 26432 & 12.5687 & 13.9 \\
& & & & 13.8351 & 6.8 \\
& & & & 10.4756 & 3.0 \\
191850	&	3860.48608 & 2246.91480 & 25537 & 13.5330 & 35.1 \\
& & & & 21.0441 & 8.8 \\
& & & & 23.3159 & 8.7 \\
& & & & 9.7829 & 6.6 \\
& & & & 24.8817 & 6.1 \\
& & & & 12.9278 & 4.2 \\
& & & & 9.6737 & 3.1 \\
& & & & 18.8933 & 3.0 \\
& & & & 36.8489 & 2.6 \\
& & & & 14.6698 & 2.0 \\
261904	&	5137.62695 & 485.73194 & 1178 & & & 1.8 \\
290492	&	4743.49561 & 1231.90771 & 24418 & & & 0.8 \\
290799	&	5496.47900 & 478.92432 & 3499 & 22.5300 & 7.2 \\
294253	&	5496.47900 & 478.92432 & 3336 & & & 1.5 \\
\hline \\ 
\multicolumn{7}{l}{$^a$ There might be frequencies in the range 1 to 5\,d$^{-1}$ present, see text.} \\
\end{tabular}   
\end{table*}

%
\section{Analysis}\label{analysis}

For eight targets, we detected variability (Table~\ref{results_data}). Three of them (HD\,101108, HD\,125889, and HD\,184779) are newly
discovered pulsators. For seven stars, no statistically significant frequencies were detected. However, we were able to lower the upper
limits compared to those previously published. In Fig.~\ref{fourier}, the Fourier spectra of HD\,83041 (single frequency), HD\,125889 (multiperiodic), 
and HD\,294092 (no significant frequency detected) are presented. It is well known that
WASP data are affected by daily aliases and systematics at low frequencies \citep{Smal11}. The noise at these frequencies (lower than 5\,d$^{-1}$),
is certainly not white but, except for HD~83277, we did not detect any suspicious peak below this limit.

Figure~\ref{instability} shows the \logTeff\ versus \logl\ diagram of our programme stars (Table~\ref{targets_basics}) together with the borders
of the $\delta$\,Scuti and $\gamma$\,Doradus instability strips taken from \citet{Breg98} and \citet{Dupr04}, respectively. 
Four stars (HD\,23392, HD\,36726, HD\,261904, and HD\,294253) are well outside the instability strips whereas HD\,90821 is just
on the edge. Variability was not detected for any of these stars, with upper limits between 0.8 and 3.0\,mmag. Two apparently constant
stars (HD\,83277 and HD\,290492) are within the instability strips. The case of HD\,83277 is discussed below.

Three stars (HD\,83041, HD\,120896, and HD\,125889) could be, with the errors of \logTeff\ and \logl, located in the 
$\gamma$\,Doradus instability strip. Recently, \citet{Paun14} discovered that the $\lambda$\,Bootis star HD\,54272
is also a $\gamma$\,Doradus type pulsator. For HD\,83041 and HD\,125889 we were not able to detect any significant
frequency in the $\gamma$\,Doradus domain (0.8 to 5\,d$^{-1}$). However, the lowest frequency (3.0111\,d$^{-1}$) found
for HD\,120896 could be due to $\gamma$\,Doradus type pulsation or due to alias effects.

In the following, we discuss the results of some targets in more detail. 

{\it HD\,83041:} \citet{Paun96} published a frequency of 15.16\,d$^{-1}$. The difference to our result (14.5293\,d$^{-1}$) can
be well explained by the very broad peak in their fourier spectrum due to the
short time basis of the earlier observations and the resultant
spectral window function.

{\it HD\,83277:} for the range of frequencies higher that 5\,d$^{-1}$, we find no significant amplitudes above 3\,mmag. This 
result is in line with those by \citet{Paun02a}. However, the frequency range between 1 to 5\,d$^{-1}$ shows a rather rich spectrum
which is just at a 4\,$\sigma$ level. Follow-up observations are needed to confirm the reality of these frequencies. 

{\it HD\,101108:} this is a multiperiodic pulsator with amplitudes below 2\,mmag. We find a rich spectrum of frequencies
in the range between 6 and 30\,d$^{-1}$. Due to the low amplitudes, we are not able to identify individual frequencies
common in all data sets. This star is a very interesting target for follow-up observations.

{\it HD\,105058:} our result (19.8097\,d$^{-1}$) is not in line with that (24.8\,d$^{-1}$, about 58\,min) published by \citet{Paun98}.
The latter observed this star for 1.8\,h covering less than two pulsation cycles. This fact, together with a possible multiperiodic
behaviour, could explain the difference.

{\it HD\,290799:} the detected frequency is exactly 1\,d$^{-1}$ lower than published by \citet{Paun02a}. They have observed this
object for 7.8\,h in Str{\"o}mgren $b$. There are two WASP datasets from which the second one shows significantly 
larger amplitudes than the first one. HD\,290799 is located in the Orion OB1 association. There are several fainter stars
in its vicinity which results in a higher noise level than for the other targets.

\begin{figure*}
\includegraphics[angle=180, width=17cm]{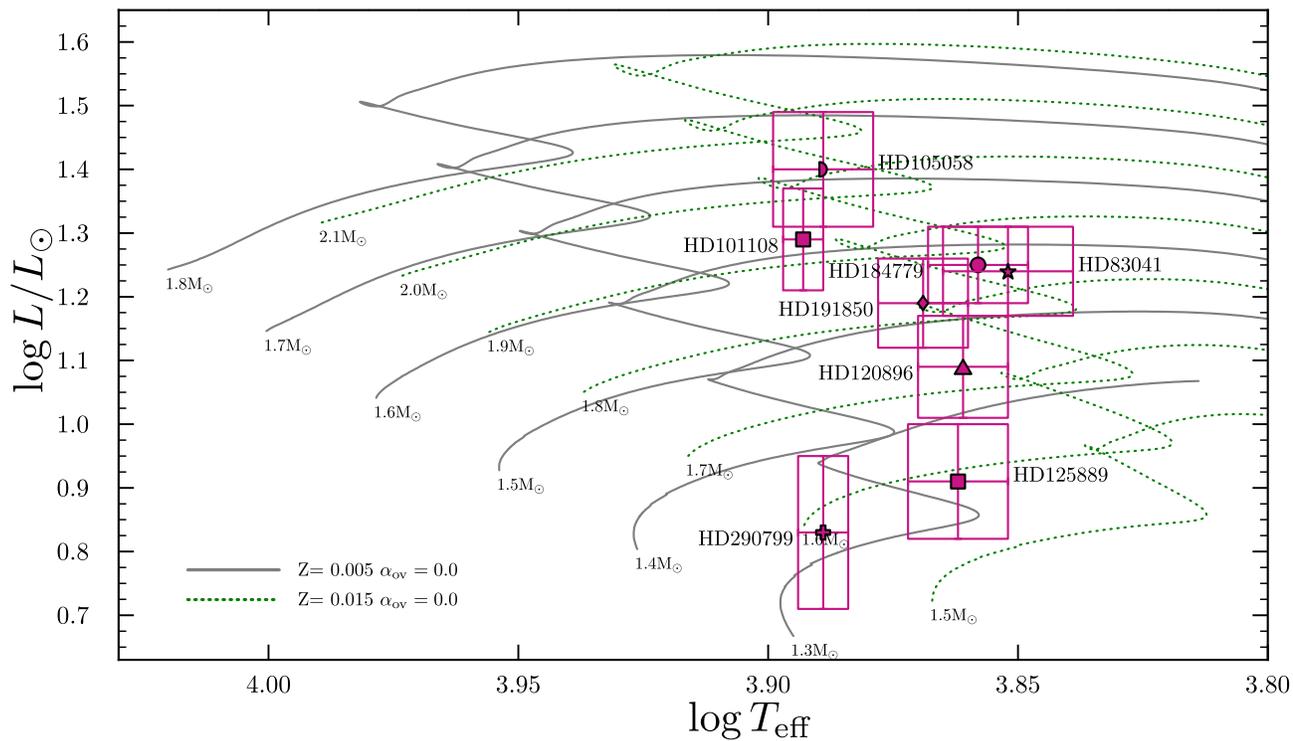}
\caption{The location of the eight variable stars (Table \ref{results_data}) within the Hertzsprung-Russell diagram.
The evolutionary tracks for two different metallicities were calculated as described in Sect. \ref{multi}. 
The initial rotational velocity was set to $V_{\rm{rot}}$\,=\,100\,km\,s$^{-1}$.}
 \label{HR1}
\end{figure*}

\begin{figure*}
\includegraphics[angle=180, width=17cm]{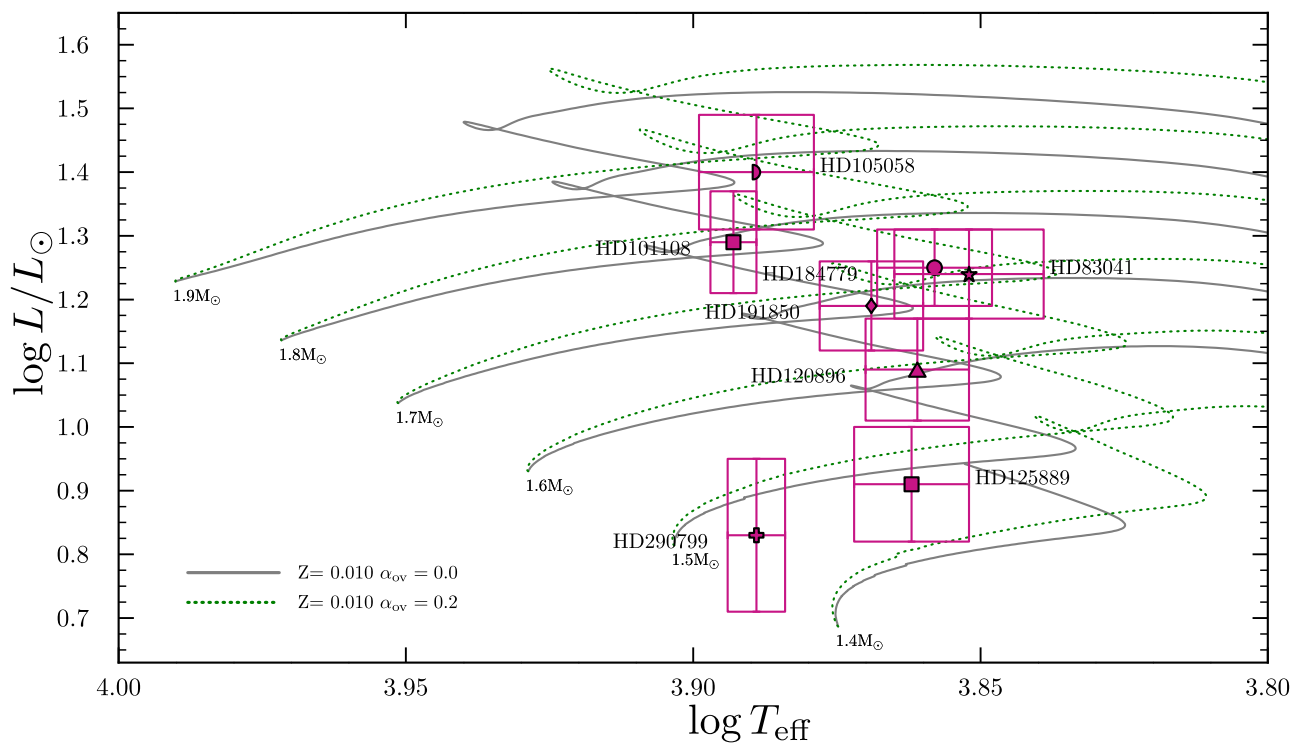}
\caption{The same as in Fig.\,\ref{HR1} but for two values of the overshooting coefficient $\alpha_{\rm{ov}}=0.0$ and 0.2, respectively.}
 \label{HR2}
\end{figure*}

\subsection{Models for the multiperiodic target stars} \label{multi}
The important question we try to answer with our data set is whether or not $\lambda$\,Bootis stars are
intrinsically metal-weak (i.e. metal-weak Pop\,I stars). This does not appear to have been raised very often in the
literature. The basic assumption for such an investigation is that all 
$\lambda$\,Bootis stars are still before the terminal-age main-sequence.
\citet{Casa09}, for example, made a detailed asteroseismologic
analysis of the pulsational behaviour of 29~Cygni (HR~7736, A0.5\,Va$^{-}$\,$\lambda$ Boo; \citealt{Gray88}). This star is one of
the prototypes of the $\lambda$\,Bootis group and exhibits very strong
underabundances of the Fe-peak elements compared to the Sun \citep{Paun02b}. From
their best-fitting models they concluded that HR~7736 is an intrinsically metal-weak
main-sequence object. We investigated the five multiperiodic target stars HD\,105058, HD\,120896, HD\,125889, HD\,184779, and HD\,191850 in that respect.

For such an analysis, the accurate knowledge of the basic stellar parameters (effective temperature, surface gravity, and metallicity)
together with the rotational velocity are needed.

First of all, we explored the \vsini\ values and metallicities ([M/H]) of these stars in more detail. 
In the ESO archive\footnote{http://archive.eso.org} FEROS spectra of these stars, except for HD\,105058, with a resolution of about 
48\,000, covering a spectral range from 3800 to 7900\AA, are available. Initial reductions of the spectra 
and their conversion into 1D images were carried out
within {\sc iraf}\footnote{http://iraf.noao.edu/}. 

The \vsini\ values were determined measuring the Full-Width at Half-Maximum (FWHM) of the Mg\,{\sc ii}\,4481\AA\ line and
comparing them with the standard relation listed by \citet{Sle75}. We compared the derived FWHM with those of other comparable
strong lines and find an excellent agreement. For HD\,120896, HD\,125889, HD\,184779, and HD\,191850, we find \vsini\ values of 125, 95, 80, and
60\,km\,s$^{-1}$, respectively. The errors are about $\pm$5\,km\,s$^{-1}$.

Synthesized spectra were computed using the program {\sc spectrum}\footnote{http://www.appstate.edu/{\raise.17ex\hbox{$\scriptstyle\mathtt{\sim}$}}grayro/spectrum/spectrum.html} 
\citep{Gray94} and modified versions of the {\sc atlas9} code taken from the Vienna New Model Grid of Stellar Atmospheres, 
{\sc nemo}\footnote{http://www.univie.ac.at/nemo} \citep{Hei02}. The astrophysical parameters were taken from Table~\ref{targets_basics}.
The spectra were convolved with the instrumental profile and the rotational profiles using the \vsini\ values
listed above. A visual comparison of the observed and synthetic line profiles yields an excellent agreement. To estimate the [M/H] value, we used 
different scaled metallicity models from +0 to $-$2\,dex. 

For HD\,120896, HD\,125889, and HD\,184779, we find a very similar metallicity of about $-$0.5\,dex ([Z]\,=\,0.004) whereas HD\,191850 is more underabundant with
[M/H]\,=\,$-$0.8\,dex ([Z]\,=\,0.002). The latter is in perfect accordance with the result of the RAdial Velocity Experiment (RAVE) survey \citep{Sie11}.

For HD\,105058 the same techniques were applied to the spectrum published by \citet{Tom10} which has a resolution of only 20\,000.
The result is \vsini\,=\,135$\pm$5\,km\,s$^{-1}$ and [M/H]\,=\,$-$1.0\,dex. 

As a next step, theoretical tracks were calculated for several different values of metallicity, 0.004\,$<$\,[Z]\,$<$\,0.025.
We used the Opacity Project data 
\citep{Seat05}, the scaled chemical composition by \citet{Aspl09}, the Warsaw-New Jersey evolutionary code 
\citep{Pamy98}, and the linear non-adiabatic pulsational code by \citet{Dzie77}.

\subsection{Results from the models}

Figure~\ref{HR1} shows the Hertzsprung-Russell diagram (HRD) with the location of all variable stars. 
The observational values of the effective temperature and luminosity were taken from Table~\ref{targets_basics}. 
>From this figure we conclude that the stars are on the main-sequence only in the case of metallicity [Z]\,$>$\,0.007 for no convective 
core overshooting, i.e. $\alpha_{\rm{ov}}$\,=\,0.0. To investigate the influence of the latter, we also calculated
evolutionary tracks for two values of the overshooting parameter, namely $\alpha_{\rm{ov}}$\,=\,0 and 0.2.
Figure~\ref{HR2} shows that the effect is quite strong. A higher efficiency of the core overshooting results 
in a much more extended main-sequence phase. If we assume $\alpha_{\rm{ov}}$\,=\,0.2 then [Z]\,$>$\,0.004 is needed
for locating the stars on the main sequence. This value is still much higher than deduced from 
the spectra. From this analysis we would conclude that these objects are intrinsically
not metal-weak.

As the next step, we performed a detailed asteroseismic analysis of the detected frequencies in the light of different
metallicity values. For this, we used the results of HD\,120896, HD\,184779, and HD\,191850 because they have
very similar astrophysical parameters.

{\it Main-Sequence hypothesis:} to place the stars on the main-sequence, we need to assume a metallicity [Z]\,$>$\,0.007 or 
include effective convective core overshooting. For $\alpha_{\rm{ov}}=0.2$ we need about [Z]\,$>$\,0.004. So, if the stars 
are on the main-sequence, we can derived some constrains on the lowest plausible value of the metallicity (Figs.~\ref{HR1} and \ref{HR2}). 

\begin{figure*}
\includegraphics[angle=180, width=17cm]{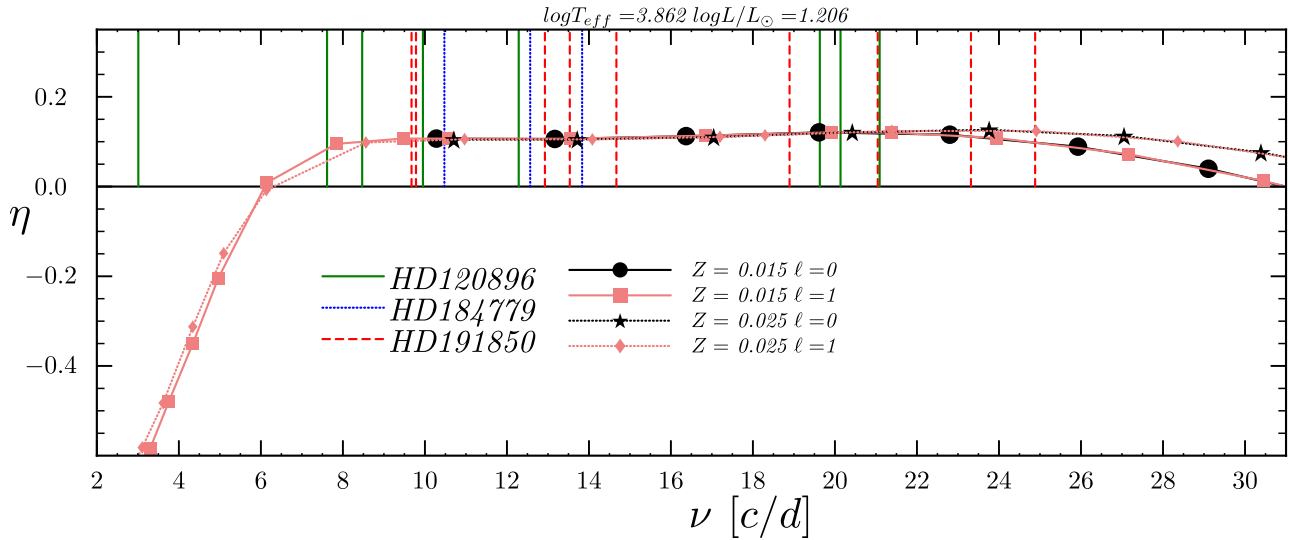}
\caption{The instability parameter, $\eta$, as a function of the pulsational frequency for three star models 
on the main-sequence. All models have \logTeff\,=\,3.862 and \logl\,=\,1.206. 
We assumed initial hydrogen abundance X\,=\,0.7 and mixing-length parameter $\alpha_{\rm{cv}}=1.8$. Different colours indicate 
various metallicities, circles indicate radial modes ($\ell$\,=\,0) and squares -- dipole modes ($\ell$\,=\,1). Short vertical lines 
above the $\eta$\,=\,0 line indicate the observational frequency spectrum for HD\,120896 (green), HD\,184779 (blue), and HD\,191850 (red).}
 \label{eta_MS}
\end{figure*}

Figure~\ref{eta_MS} shows the instability parameter $\eta$, as a function of the frequency for pulsational models. Modes are excited when $\eta$\,$>$\,0.
The models have \logTeff\,=\,3.862 and \logl\,=\,1.206. Different colours indicate 
various metallicities, circles indicate radial modes ($\ell$\,=\,0) and squares -- dipole modes ($\ell$\,=\,1). Short vertical lines 
above the $\eta$\,=\,0 line indicate the observational frequency spectrum for HD\,120896 (green), HD\,184779 (blue), and HD\,191850 (red).
We notice that the unstable modes cover almost the entire observational spectrum. Increasing the initial hydrogen abundance 
does not change the results significantly. The same is true if we change the mixing-length parameter to $\alpha_{\rm{cv}}$\,=\,1. 
A more significant impact on the models is the overshooting efficiency. For $\alpha_{\rm{ov}}$\,=\,0.2, we were able to derive models 
for metallicity [Z]\,=\,0.005, while for $\alpha_{\rm{ov}}$\,=\,0.0, we need [Z]\,=\,0.008.

We conclude that if the stars are on the main-sequence, a much higher intrinsic metallicity than the values derived from observations
are needed to explain the observed frequencies. However, only a detailed mode identification, which is not possible from the presented 
data, would allow to define even more strict constrains for the models. 

\begin{figure*}
\includegraphics[angle=180, width=17cm]{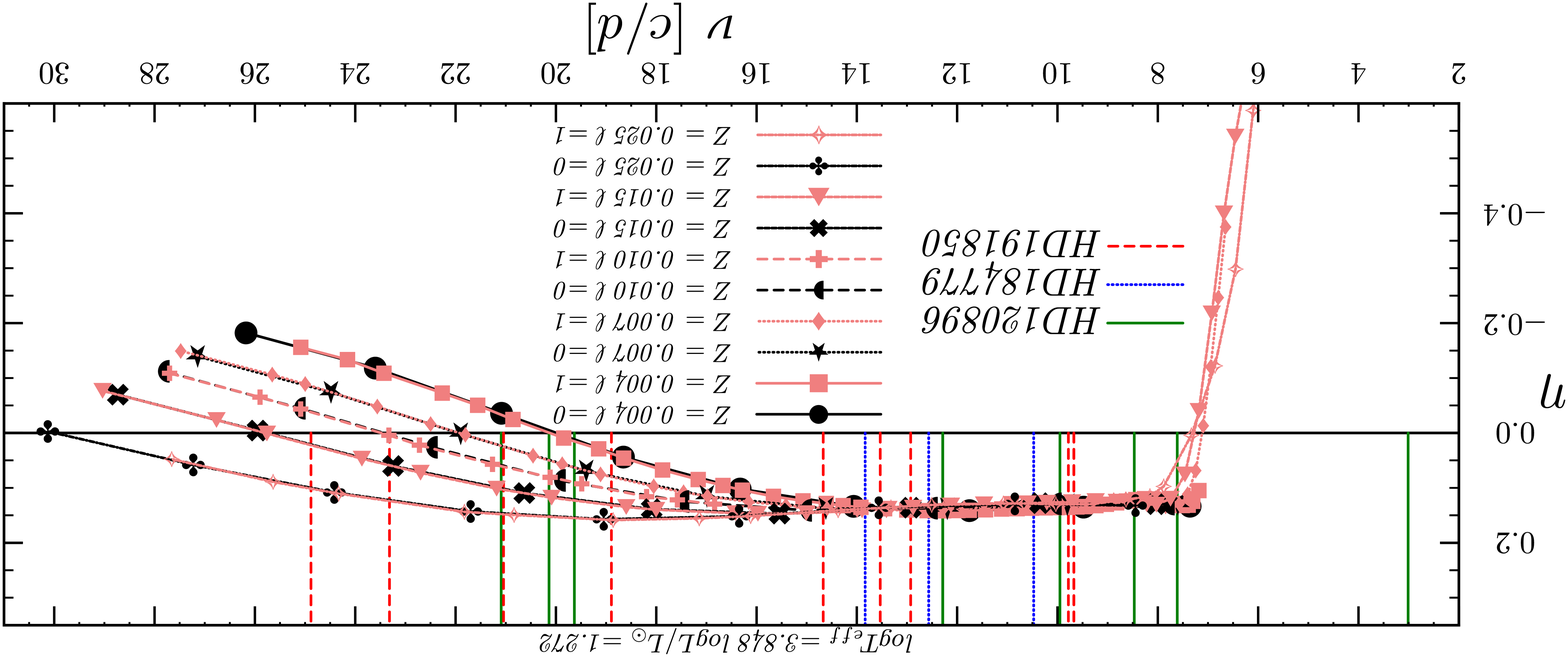}
\caption{Same as Fig. \ref{eta_MS} but for \logTeff\,=\,3.848 and \logl\,=\,1.272.}
 \label{eta_HB}
\end{figure*}  

{\it Post Main-Sequence hypothesis:} in Fig.~\ref{eta_HB} we present the same analysis as in Fig.~\ref{eta_MS} but for
four different metallicities, [Z]\,=\,0.004 (black), 0.007 (green), 0.015 (blue) and 0.025 (red). The left instability border appears at a frequency 
that is sufficiently low to explain almost all observed modes. The only exception is a very low-frequency of HD\,120896. This mode cannot be 
the retrograde mode shifted to low-frequency due to the fast rotation; it would require a rotational velocity of the order of 1000\,km\,s$^{-1}$.
The low-frequency instability border at about $\nu$\,$\sim$\,7\,d$^{-1}$ is almost insensitive to the metallicity, but the high-frequency border 
depends strongly on [Z]. Because the interaction of convection and pulsation are not well described in the present theory, the true 
position of the high-frequency instability border is quite uncertain. 

A higher value of the initial hydrogen content slightly increases the instability of high-frequency modes. It is due to the fact that 
high-frequency modes are partially driven through the H\,{\sc{i}}-ionization zone. High-frequency modes have an increase of the work integral near 
the H\,{\sc{i}} opacity bump. Since convection transports almost the entire energy 
in this zone, this pulsation driving effect can be artificial.
For the chosen effective temperature and luminosity, we were not able to find models beyond the main-sequence for [Z]\,$>$\,0.025.
For a more effective core overshooting, we could not find models with [Z]\,$>$\,0.015. The tracks are shifted to the right on the 
HRD, and the loop after the main-sequence appears for lower values of the effective temperature. The mixing-length parameter, $\alpha_{\rm{cv}}$, 
was set to 1.8. The change of this parameter to 1.0 did not cause a significant effect.

In this regime, there are quite strong effects of the chosen $T_{\rm{eff}}$ and \logl\ values. The instability of modes depends 
quite strongly on the exact position of a star on the HRD. For a given [Z], a higher effective temperature gives instability for higher frequencies. 
The low-frequency instability border is shifted to about 9\,d$^{-1}$. For cooler models, we derived instability for lower frequencies, 
but high-frequency modes were slightly more stable. More luminous models could not explain the instability of high-frequency modes, while less luminous ones
have problems with low-frequency modes. This is interesting because the lowest frequencies were found for HD\,120896, which is the least luminous star. 
But its effective temperature is also lower than for the remaining stars, and this effect should cancel the luminosity effect.

The theoretical evolutionary changes of frequencies are of order of $10^{-7}$\,d$^{-1}$\,yr$^{-1}$, which are too small to detect with the current data.

We conclude that with the frequencies found in this work we are not able to reject the possibility that the stars are beyond the main-sequence. Both
low and high metallicities are possible because the instability of modes depends mainly on the helium ionization zone. 

\section{Conclusion}\label{conclusion}
We analysed the time series of WASP datasets of 15 well established $\lambda$\,Bootis stars.
This small group of chemically peculiar objects of the upper main-sequence is an excellent
target to investigate the effects of diffusion, rotation, and accretion in the presence of
classical $\delta$\,Scuti/$\gamma$\,Doradus type pulsation.

For eight targets, we were able to detect (multiperiodic) variations with amplitudes between
1.6 and 35.1\,mmag. Four of the probable constant stars are not located in the instability 
strip. A comparison of our results with those from the literature yields an excellent agreement.

We made an asteroseismic analysis of the multiperiodic stars to tackle the question
as to whether the chemical peculiarity is intrinsic or restricted to the stellar surface. For this,
we estimated the metallicities and projected rotational velocities from high-resolution archival
spectra. We then used state-of-the-art pulsation models to characterize the (in)stability
properties. From this analysis we conclude that if the stars are still on the main-sequence, which
is the most accepted hypothesis \citep{Stu06}, they are not intrinsically metal-weak. If they are beyond the
terminal-age main-sequence, the results are ambiguous. We suggest detailed spectroscopic follow-up 
observations of the presented multiperiodic $\lambda$\,Bootis stars to identify individual modes.
Such spectroscopic observations have been already successfully performed for one member of the group, 29~Cygni, by \citet{Mkrt07}.
Similar studies for $\delta$ Scuti stars of comparable magnitudes have been published \citep{Pore09} proving the feasibility.
The expected high precision parallaxes and thus distances as well as luminosities from the Gaia satellite
mission \citep{Mich15} will hopefully significantly improve the accuracy of the position of the investigated stars
in the HRD. At least for those which are not too bright.
\section*{Acknowledgements}
The WASP project is funded and maintained
by Queen's University Belfast, the Universities of Keele, St.
Andrews, Warwick and Leicester, the Open University, the Isaac
Newton Group, the Instituto de Astrofisica Canarias, the South
African Astronomical Observatory and by the STFC. This project was supported by the SoMoPro II Programme (3SGA5916),
co-financed by the European Union and the South Moravian Region, the 
grant GA \v{C}R 7AMB12AT003, LH14300, and
the financial contributions of the Austrian Agency for International 
Cooperation in Education and Research (BG-03/2013 and CZ-09/2014).
PW was supported by the Polish National Science Centre grants No DEC-2013/08/S/ST9/00583.
Calculations have been partly carried out using resources provided by Wroclaw Centre 
for Networking and Supercomputing (http://www.wcss.pl), grant No. 265.
We made use of ``The catalog of MK Spectral Types'' by B. Skiff.
We thank the anonymous referee for the careful reading of our manuscript and the 
many insightful comments and suggestions.
This work reflects the opinion of the authors and the European 
Union is not responsible for any possible application of the information
included in the paper.

\label{lastpage}
\end{document}